\documentclass[structabstract]{aa} 
\usepackage{longtable,booktabs}
\usepackage{latexsym}
\usepackage{amssymb}
\usepackage{lscape}
\usepackage[]{natbib}
\usepackage{subfigure}
\usepackage{xcolor}
\usepackage{multicol}
\usepackage{multirow}
\usepackage{amsmath}
\usepackage{booktabs}
\usepackage{caption}
\usepackage{tablefootnote}
\usepackage{pdflscape}
\usepackage{graphicx}
\usepackage{txfonts}
\def\lsim{\mathrel{\rlap{\lower 3pt \hbox{$\sim$}} \raise 2.0pt \hbox{$<$}}}
\def\gsim{\mathrel{\rlap{\lower 3pt \hbox{$\sim$}} \raise 2.0pt \hbox{$>$}}}

\title{The central engine of the highest redshift blazar}

\author{S. Belladitta\inst{1,2}                               
\and A. Caccianiga\inst{1}
\and A. Diana\inst{1,3}
\and A. Moretti\inst{1}  
\and P. Severgnini\inst{1}
\and M. Pedani\inst{4}
\and \\ L. P. Cassar\`a\inst{5}
\and C. Spingola\inst{6}
\and L. Ighina\inst{1,2}
\and A. Rossi\inst{7}
\and R. Della Ceca\inst{1}
}

\institute{INAF $-$ Osservatorio Astronomico di Brera, via Brera, 28, 20121 Milano, Italy\\
\email {silvia.belladitta@inaf.it}
\and
DiSAT $-$ Universit\`a degli Studi dell'Insubria, Via Valleggio 11, 22100 Como, Italy
\and 
Dipartimento di Fisica G. Occhialini $-$ Universit\`a degli Studi di Milano Bicocca, Piazza della Scienza 3, 20126 Milano, Italy
\and
INAF $-$ Fundaci\'on Galileo Galilei, Rambla Jos\'e Ana Fernandez P\'erez 7, 38712 Bre\~{n}a Baja, TF, Spain
\and
INAF $-$ Istituto di Astrofisica Spaziale e Fisica Cosmica (IASF), Via A. Corti 12, 20133 Milano
\and
INAF $-$ Istituto di Radioastronomia Via Gobetti 101 40129 Bologna, Italy
\and
INAF $-$ Osservatorio di Astrofisica e Scienza dello Spazio, via Piero Gobetti 93/3, 40129 Bologna, Italy
}

\begin{document}
	
\date{Received; accepted}

\abstract
{We report on a LUCI/Large Binocular Telescope near-infrared (NIR) spectrum of PSO~J030947.49+271757.31 (hereafter PSO~J0309+27), the highest redshift blazar known to date (z$\sim$6.1).  
From the C$\rm IV$$\lambda$1549 broad emission line we found that PSO~J0309+27 is powered by a 1.45$^{+1.89}_{-0.85}$$\times$10$^9$M$_{\odot}$ supermassive black hole (SMBH) with a bolometric luminosity of $\sim$8$\times$10$^{46}$~erg~s$^{-1}$ and an Eddington ratio 
equal to 0.44$^{+0.78}_{-0.35}$.  
We also obtained new photometric observations with the Telescopio Nazionale Galileo in $J$ and $K$ bands to better constrain the NIR Spectral Energy Distribution of the source.   
Thanks to these observations, we were able to model the accretion disk and to derive an independent estimate of the black hole mass of PSO~J0309+27, confirming the value inferred from the virial technique.
The existence of such a massive SMBH just $\sim$900 million years after the Big Bang challenges models of the earliest SMBH growth, especially if jetted Active Galactic Nuclei are associated to a highly spinning black hole as currently thought. 
Indeed, in a Eddington-limited accretion scenario and assuming a radiative efficiency of 0.3, typical of a fast rotating SMBH, a seed black hole of more than 10$^6$ M$_{\odot}$ at z = 30 is required to reproduce the mass of PSO~J0309+27 at redshift 6. 
This requirement suggests either earlier periods of rapid black hole growth with super-Eddington accretion and/or that only part of the released gravitational energy goes to heat the accretion disk and feed the black hole.}

\maketitle
%

\section{Introduction}
\label{intro}
High redshift (z$>$6) Active Galactic Nuclei (AGNs) are direct probes of the Universe less than 1~Gyr after the Big Bang.
These earliest AGNs are fundamental to study the early growth of supermassive black holes (SMBHs, e.g. see Inayoshi et al. 2020, for a recent review).
An accurate determination of SMBH masses of z$>$6 AGNs is a prerequisite to fully understand SMBHs physics, demographics (e.g. black holes mass function), and relations with  their host galaxies.
With the so-called single-epoch (SE) method (e.g., Vestergaard \& Peterson 2006; Vestergaard \& Osmer 2009; Shen et al. 2008, 2011, 2019; Trakhtenbrot \& Netzer 2012) it has been possible to estimate the mass of the SMBHs hosted by high-z AGNs (e.g., Jiang et al. 2007; Kurk et al. 2007; Wu et al. 2015; Mazzucchelli et al. 2017; Kim et al. 2018; Shen et al. 2019; Onoue et al. 2019; Yang et al. 2020; Andika et al. 2020; Wang et al. 2021).
These studies have shown that high-redshift AGNs discovered to date are typically powered by SMBHs more massive than 10$^8$-10$^9$ M$_{\odot}$, comparable to the most massive black holes at any redshift.
These discoveries indicate a fast and efficient growth of black holes that challenges the currently accepted theoretical model of SMBHs formation (e.g. Volonteri 2010, Latif \& Ferrara 2016).
The most popular scenarios to explain the mass assembly of several million solar mass black holes in the early Universe include direct collapse of massive gas clouds (e.g., Haehnelt \& Rees 1993; Begelman et al. 2006; Latif \& Schleicher 2015), the collapse of Population III stars (e.g., Bond 1984; Alvarez et al. 2009; Valiante et al. 2016), the co-action of dynamical processes, gas collapse and star formation (e.g., Devecchi \& Volonteri 2009), or intense gas accretion in a super-Eddington phase (e.g., Alexander \& Natarajan 2014; Madau et al. 2014; Lupi et al. 2016; Pezzulli et al. 2016; Volonteri et al. 2016). \\
Even more challenging is the discovery of high-z massive SMBHs hosted in radio-loud\footnote{Here we consider an AGN to be RL if it has a radio loudness (R) larger than 10, with R defined as the ratio between the 5~GHz and 4400$\mbox{\AA}$ rest frame flux densities, R = $\frac{S_{5GHz}}{S_{4400\AA}}$ (Kellermann et al. 1989).} (RL or jetted) AGNs. 
RL AGNs are characterized by the presence of two collimated relativistic jets of plasma emitted from the central SMBH and extended up to a few Mpc (e.g., see Blandford et al. 2019 for a recent review).
The presence of relativistic jets is usually associated to a highly spinning accreting black hole (e.g., Blandford \& Znajek 1977; Tchekhovskoy et al. 2011), which is expected to have a large radiation efficiency ($\eta \sim$0.3; e.g., Thorne et al. 1974) and, therefore, for a given luminosity, a longer growth time with respect to black holes hosted by radio-quiet (RQ or non-jetted) AGNs. 
Hence, since there is not enough time to accrete large masses (M$_{\rm_{BH}}$ $\sim10^9$ M$_{\odot}$) in a standard Eddington-limited accretion scenario, super-Eddington accretion events can be invoked to explain the existence of these high-z jetted SMBHs (e.g., Begelman \& Volonteri 2017; Yang et al. 2020). Therefore, identifying and characterizing high-z RL AGNs provide a unique opportunity to study the role of jets in the accretion of SMBHs in the primordial Universe (e.g., Volonteri et al. 2015). \\
If a RL AGN has its relativistic jets oriented along the line of sight, we classify it as a \textit{blazar} (e.g., Urry \& Padovani 1995; Padovani et al. 2017). 
Since the jet emission is strongly boosted and not obscured along the jet direction, the observed luminosity of blazars is usually very high, making these sources well visible up to very high-z.
Although blazars represent a small fraction of RL AGNs, they are fundamental to ensure a reliable and complete census of the global population of the jetted AGNs and, therefore, to trace the evolution of the SMBHs across the cosmic time (e.g., Ajello et al. 2009; Ghisellini et al. 2010b; Sbarrato et al. 2015; Caccianiga et al. 2019, Ighina et al. 2021b; Diana et al. 2021). 
Indeed, from the space density of blazars it is possible to infer the space density of all the RL AGNs that have similar intrinsic physical properties.
If we define as blazar a source observed within an angle equal to 1/$\Gamma$, where $\Gamma$ is the bulk Lorentz factor of the emitting plasma, we expect to find N$_{RL~AGNs}$ = N$_{blazars}$ $\times$ 2$\Gamma^2$, (e.g., Volonteri et al. 2011). 
Therefore, the discovery of high-z blazars ensures the census, free from obscuration effects, of early SMBHs and provides strong and critical constraints on the accretion mode, the mass and the spin of the first seed black holes (e.g., Kellerman 2016).\\
Recently, we discovered the most distant blazar to date, PSO~J030947.49+271757.31 (hereafter PSO~J0309+27 at z$\sim$6.1; Belladitta et al. 2020). 
We are carrying out a multi-wavelength study on this source, from the radio (Spingola et al. 2020) to the X--ray band (Moretti et al. 2021, Ighina et al. 2022), in order to characterize the properties of this very distant jetted AGN at all wavelengths.
Here we present new observations in the near-infrared (NIR) band. They consist on a Large Binocular Telescope (LBT) spectroscopic observation carried out to detect the C$\rm IV$$\lambda1549$ (hereafter C$\rm IV$) emission line useful for the computation of the central black hole mass, and on photometric observations in $J$ and $K'$ bands obtained at the Telescopio Nazionale Galileo (TNG) to better constrain the Spectral Energy Distribution (SED) of the source.\\ 
The paper is structured as follows: in Section \ref{allobs} we present the LBT and TNG NIR observations of PSO~J0309+27; in Section \ref{resdisc} we report the data analysis and the results of our observations (i.e., C$\rm IV$ line characterization, NIR magnitudes, black hole mass estimations); in Section \ref{psogrowth} we discuss our results on the black hole mass in term of seed black hole growth; the conclusions are reported in Section \ref{conc}.\\
Throughout the paper we use a flat $\Lambda$CDM cosmology, with H$_{0}$ = 70 km s$^{-1}$ Mpc$^{-1}$, $\Omega_m$ = 0.3 and $\Omega_{\Lambda}$ = 0.7. All errors are reported at 1$\sigma$, unless otherwise specified.

\section{Spectroscopic and photometric observations}
\label{allobs}
\subsection{LBT/LUCI}
\label{lbtobs}
\begin{figure*}[!h]
	\centering
	\includegraphics[width=13.5cm]{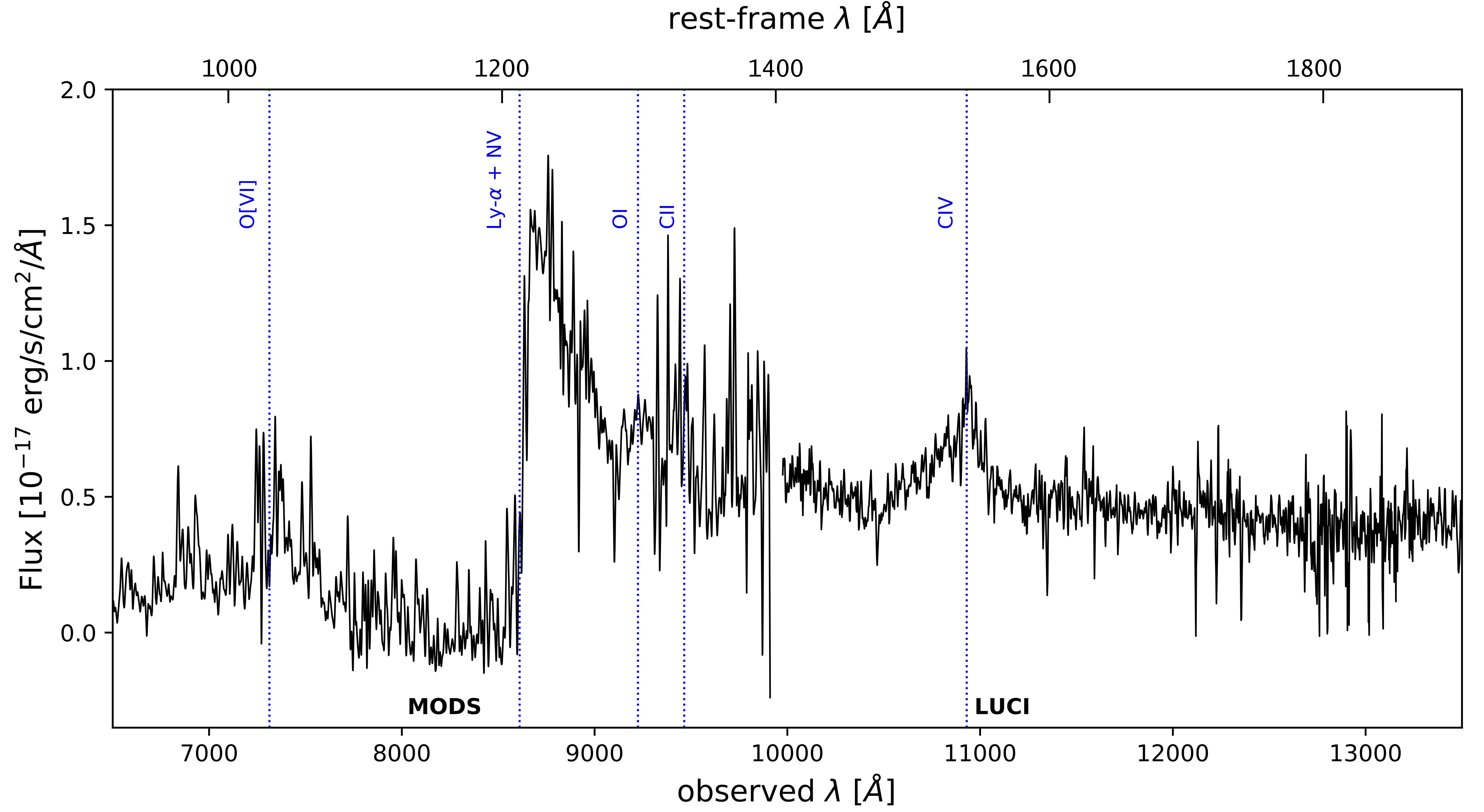}
	\vskip -0.3 true cm
	\caption{\small LBT MODS and LUCI observed spectra of PSO~J0309+27. Together with the optical lines already marked in the MODS spectrum in Belladitta et al. (2020), here the C$\rm{IV}$$\lambda1549$ line is marked. 
	On the upper x-axis the rest frame wavelengths are shown. }
	\label{spec}
\end{figure*}

The high-z nature of PSO~J0309+27 (z$\sim$6.1) prevents the detection in the optical wavelength range of the C$\rm{IV}$$\lambda1549$ and/or the Mg$\rm{II}$$\lambda2798$ broad emission lines typically used for the black hole mass estimation. 
The latter, at the redshift of the object, falls in an atmospheric absorption band, hence it is not easily detectable by ground based telescopes.
Therefore, we proposed an LBT Utility Camera in the Infrared (LUCI, Seifert et al. 2003) follow-up in order to extend the wavelength range in the NIR band to detect the C$\rm{IV}$ broad emission line.
The observation was carried out in a Director's Discretionary Time program (program ID: DDT\_2019B\_3; PI: S. Belladitta) on 2019 December 2 and consisted of 12 exposures of 15 minutes each, with a
long-slit of 1.2$''$ width, in nodding mode in the sequence ABBA, with a total integration time of 3 hours. 
The medium seeing throughout the night was 1.1$''$ and the mean air mass was 1.2.
We used the G200-zJ configuration for both LUCI1 and LUCI2, in order to cover the wavelength range from 0.9 to 1.2 $\mu$m, where the C$\rm IV$ was expected to be found. 
The data reduction was performed at the Italian LBT Spectroscopic Reduction Center.
Each spectral image was independently dark subtracted and flat-field corrected. Sky subtraction was done on 2D extracted, wavelength calibrated spectra.
Wavelength calibration was obtained by using several sky lines, reaching a rms of 0.33$\mbox{\AA}$ on LUCI1 and of 0.25$\mbox{\AA}$ on LUCI2.
The LBT/LUCI spectrum of PSO~J0309+27 is shown in Fig.~\ref{spec}, together with the LBT/MODS spectrum already reported in Belladitta et al. (2020).
Both the spectrum and the photometric points have been corrected for Galactic extinction, using the extinction law provided by Fitzpatrick (1999), with a R${_V}$ = 3.1.
The C$\rm{IV}$ emission line is clearly detected, and it can be used to compute the mass of the central SMBH of PSO~J0309+27.

\subsection{TNG/NICS}
\label{tngobs}
\begin{figure}[!h]
	\centering
	{\includegraphics[width=5.0cm, height=5.0cm]{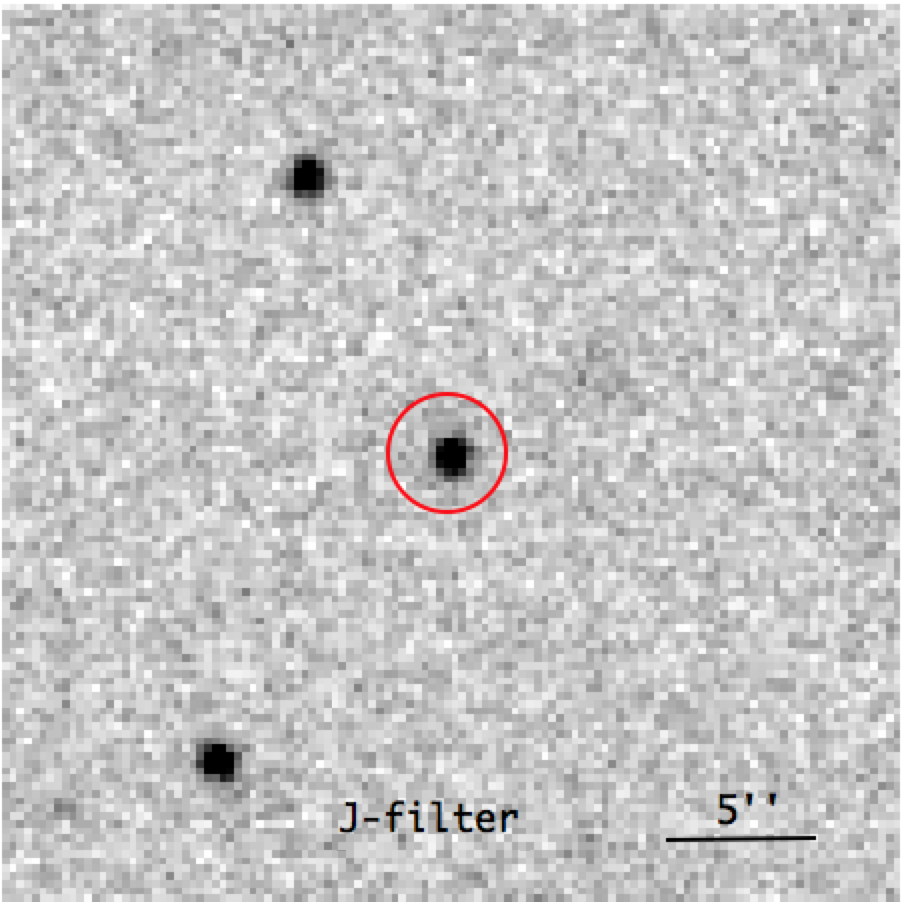}\hspace{0.3mm}
		\includegraphics[width=5.0cm, height=5.0cm]{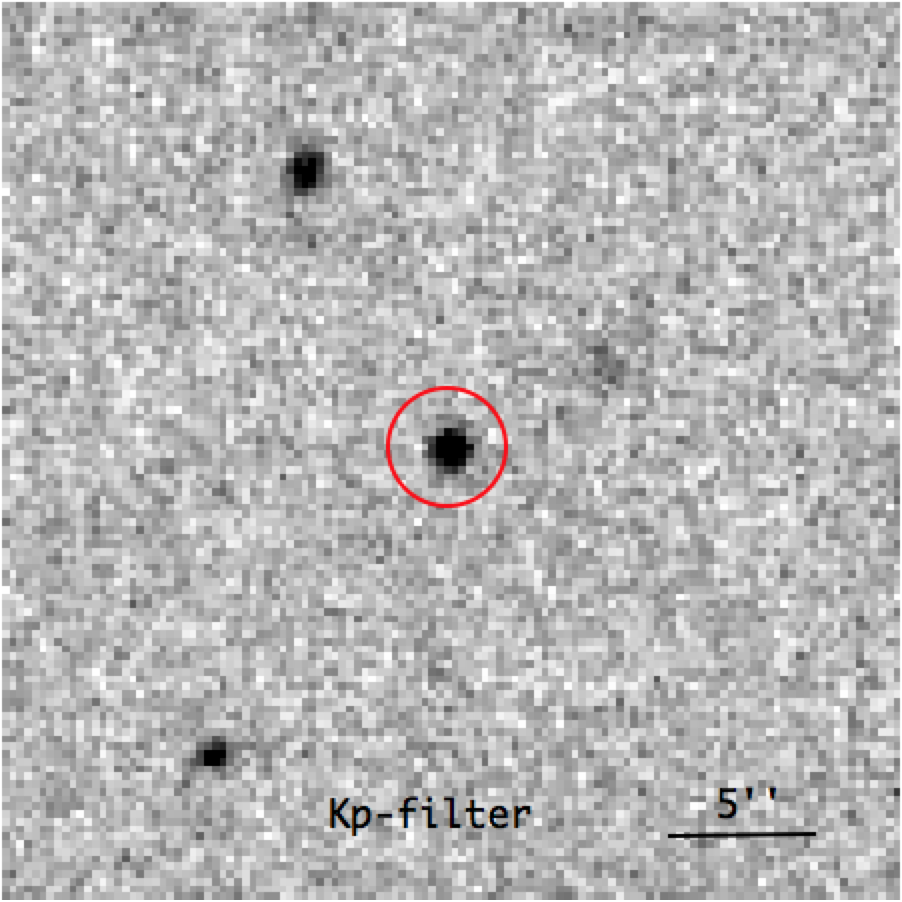}}
	\caption{\small 0.5$'$$\times$0.5$'$ $J$ and $K'$ cutout images of PSO~J0309+27 taken with the LF camera mounted on the NICS instrument. Its optical position is marked with a red circle of 2$''$ of diameter. The two images are oriented with north up and east to the left. }
	\label{tngimages}
\end{figure}

PSO~J0309+27 was observed with $J$ and $K'$ filters (central $\lambda$ = 1.27 $\mu$m and 2.12 $\mu$m respectively) at the TNG during the night of February 12, 2021 (program ID: A42DDT4, PI: S. Belladitta) with the large field camera mounted on the Near Infrared Camera Spectrometer (NICS) instrument (Baffa et al. 2001) under excellent seeing conditions (FWHM$\sim$0.75$''$, see Table \ref{tngtable}). 
The observations consisted of 50-positions dithered-mosaic with a DIT of 1$\times$60s ($J$-band) and 3$\times$20s ($K'$-band) with a total integration time of 50 minutes per band\footnote{Actually the final integration time in $J$ band is 2940~s since one exposure was lost.}.
Flux calibration has been ensured by a short (30s) observation of the field of the AS13\footnote{There are four photometric stars in this field: AS13-0, AS13-1, AS13-2, AS13-3. However the AS13-0 is saturated, hence the photometric ZP was computed by using the remaining three.} (RA = 05:57:07.5, DEC = 00:01:11) photometric standard stars (ARNICA catalog; Hunt et al. 1998), just after the object acquisition. 
Standard data reduction, such as flat fielding, sky-subraction, cross-talk effect, image alignment and stacking, was performed with the Speedy Near-IR data Automatic reduction Pipeline (SNAP\footnote{\url{http://www.tng.iac.es/news/2002/09/10/snap/index.html}}) properly optimized for NICS data. 
$J$ and $K'$ images of PSO~J0309+27 are reported in Fig. \ref{tngimages}. The source appears point-like in both bands.
The AB zero-point (ZP) magnitudes are reported in Table \ref{tngtable}, along with the total exposure time and the seeing in the final mosaic, computed measuring the FWHM of reliable point-like objects by using the Image Reduction and Analysis Facility (IRAF)\footnote{\url{http://ast.noao.edu/data/software}} task $imexamine$.

\begin{small}
	\begin{table}[!h]
		\caption{\small Details on TNG observations in $J$ and $K'$ bands and PSO~J0309+27 measured magnitudes.}
		\label{tngtable}
		\centering
		\begin{tabular}{ccccc}
			\hline\hline
			Filter & $\lambda_{central}$ & ZP  & seeing & mag \\
			       & $\mu$m & AB & arcsec & AB \\
			(1) & (2) & (3) & (4) & (5) \\
			\hline
			$J$ & 1.27 & 23.25 & 0.76 & 20.81 $\pm$ 0.06  \\
			$K'$ & 2.12 & 23.92 & 0.75 & 20.93 $\pm$ 0.08 \\
			\hline
		\end{tabular}
		\tablefoot{Col (1): NICS filter; Col (2): filter central wavelength; Col (3): photometric ZP in AB system; Col(4): Seeing (FWHM); Col(5): $J$ and $K'$ AB magnitudes of the object. The relations to convert from Vega to AB systems are: $J_{AB}$ = $J_{Vega}$ + 0.91 and $K'_{AB}$ = $K'_{Vega}$ + 1.85.}
	\end{table}
\end{small}

\section{Results and discussion} 
\label{resdisc}

\subsection{C$\rm IV$ line width and luminosity}
\label{linewidth}
\begin{figure}[!h]
	\centering
	\includegraphics[width=9.2cm, height=7.2cm]{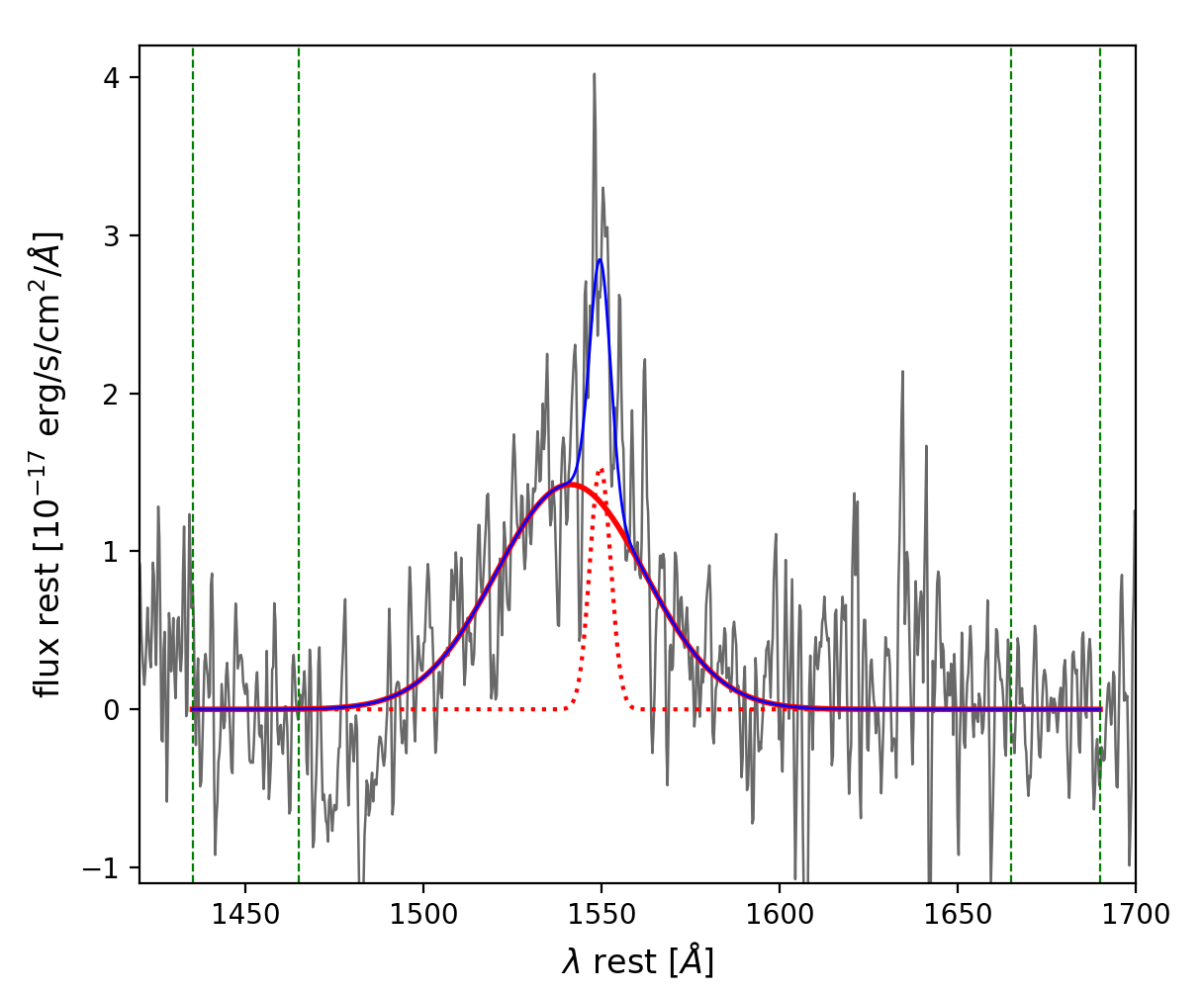}
	\caption{\small Double Gaussian fit of the C$\rm IV$ emission line. The rest frame LBT/LUCI continuum-subtracted spectrum is reported in grey, the two Gaussian functions in red and the sum of the two in blue. The larger Gaussian (solid red line; $\lambda$ $\simeq$ 1541$\mbox{\AA}$) has a FWHM of $\sim$9520 km~s$^{-1}$; instead the narrower component (dotted red line; $\lambda$ $\simeq$ 1549$\mbox{\AA}$) has a FWHM of $\sim$1330 km~s$^{-1}$.
	The intervals for the continuum selection are indicated with dashed green vertical lines.}
	\label{fit}
\end{figure}

We characterized the line width with both the Full Width at Half Maximum (FWHM) and the line dispersion ($\sigma_{line}$, as defined in Peterson et al. 2004). 
We computed both by fitting the line profile, following different steps. 
First of all we de-redshifted the LBT/LUCI spectrum 
using a redshift of 6.063$\pm$0.003, which is based on the position of peak of the C$\rm IV$ emission line (z$_{\rm CIV}$). 
Then we linearly\footnote{AGNs usually show a power-law spectrum, but in short wavelenghts intervals the linear fit is a good approximation.} fitted the continuum near the CIV line in two specific intervals (1445-1465$\mbox{\AA}$ and 1670-1690$\mbox{\AA}$, see Fig.~\ref{fit}) free from spectral features and spikes due to the background. 
On the pseudo-continuum subtracted spectrum we fitted the C$\rm IV$ broad emission line. 
Since the existence of a strong narrow component (produced by the Narrow Line Region) of the C$\rm IV$ line is controversial and difficult to detect (e.g., Wills et al. 1993; Corbin \& Boroson 1996; Vestergaard 2002; Shen \& Liu 2012), we did not include this component in the line fit.
Moreover we decided not to include the Fe$\rm II$ features, because, as mentioned in previous studies (e.g., Shen et al. 2011; Trakhtenbrot \& Netzer 2012; Zuo et al. 2020), the contribution from Fe$\rm II$ around the C$\rm IV$ line is expected to be small.
From Fig.~\ref{fit} it is clear that two Gaussian functions are necessary to reproduce the broad emission line profile properly. 
Indeed, it has been already demonstrated in several works (e.g., Laor et al. 1994; Shen et al. 2008; Tang et al. 2012) that the C$\rm IV$ broad line is usually well described by a multiple Gaussian profile and not by a single Gaussian function. 
The two Gaussian components shown in Fig. \ref{fit} have the following characteristics: the larger Gaussian is centered at $\simeq$ 1541$\mbox{\AA}$ and has a FWHM of $\sim$ 9520 km s$^{-1}$; the narrower component, at $\lambda \simeq$ 1549 $\mbox{\AA}$, has a FWHM of $\sim$ 1330 km s$^{-1}$.
In Table \ref{param} we reported the best fit parameters for the total C$\rm IV$ emission line (i.e., those derived from the sum of the two Gaussian components). Besides the FWHM (15.3 $\mbox{\AA}$) and the $\sigma_{line}$ (19.7 $\mbox{\AA}$), from the best fit model we also measured the rest-frame equivalent width (REW) and the line luminosity (L$_{\rm CIV}$ = 4$\pi$D$_{\rm L}^{2}F_{\rm CIV}$). 
The uncertainties on these values were evaluated through a Monte Carlo method (e.g., Shen et al. 2011; Shen \& Liu 2012; Raiteri et al. 2020; Zuo et al. 2020, Diana et al. 2021).
Each wavelength of the best fit model was randomly perturbed for 1000 times, according to a Gaussian distribution of the mean rms of the spectra computed underneath the C$\rm IV$ line on the pseudo-continuum subtracted spectrum. 
In this way we obtained 1000 different mock spectra of the line profile, from which we measured the line properties with the same procedure used on the real data.
We computed the distributions of FWHM, $\sigma_{line}$, REW and F$_{\rm CIV}$ for these 1000 simulated spectra, and the interval that contains 68\% of the data in these distributions was taken as the statistical uncertainty on the best fit values. 

\begin{small}
	\begin{table*}[!h]
	\captionsetup{width=.8\linewidth}
		\caption{\small Best fit parameters for the total C$\rm IV$ broad emission line.} 
	\label{param}
	\centering
	\begin{tabular}{ccccccc}
		\hline\hline
		FWHM & $\sigma_{line}$ & REW & F$_{\rm CIV}$ & L$_{\rm CIV}$ & $\lambda_{half}$ & $\Delta \rm v$ \\
		km~s$^{-1}$ & km~s$^{-1}$ & $\mbox{\AA}$ & 10$^{-16}$ erg~s$^{-1}$cm$^{-2}$ & 10$^{44}$ erg~s$^{-1}$ & $\mbox{\AA}$ &  km~s$^{-1}$ \\
		(1) & (2) & (3) & (4) & (5) & (6) & (7) \\ 
		\hline  
		 2960$^{+1030}_{-760}$  & 3815$^{+190}_{-165}$  
		 &  25.3$^{+0.7}_{-0.9}$  & 8.56$^{+0.24}_{-0.34}$ & 3.54$^{+0.10}_{-0.12}$ & 1542.16$^{+0.98}_{-0.72}$  & 1420$^{+140}_{-190}$ \\
		\hline
	\end{tabular}
	\tablefoot{Col (1) and (2): line width in term of FWHM and $\sigma_{line}$; Col (3): rest-frame equivalent width; Col (4) and (5): line flux and line luminosity; Col (6): line centroid; Col (7): line blueshift.}
\end{table*}
\end{small}

\subsection{C$\rm IV$ line blueshift and asymmetry}
\label{shift}
The C$\rm IV$ emission line is known to show asymmetry and to be blueshifted with respect to low ionization lines (e.g., Gaskell 1982; Richards et al. 2011; Coatman et al. 2017; Vietri et al. 2018; Zuo et al. 2020), independently from the source orientation (e.g., Kimball et al. 2011; Runnoe et al. 2014).
These characteristics suggest that the C$\rm IV$ clouds are affected by non gravitational effects, such as outflows, most likely originated in disk winds.  
Large C$\rm IV$ blueshifts indicate that non-virial motions have a significant effect on the observed emission velocity profile. 
To date, the largest blueshifts (>3000 km s$^{-1}$) have been discovered in the so-called weak emission line quasars (WELQs, Diamond-Stanic et al. 2009), which exhibit a REW < 10$\mbox{\AA}$ and a strongly asymmetric line profile (see e.g., Vietri et al. 2018 and reference therein). 
With a REW of $\sim$25$\mbox{\AA}$ PSO~J0309+27 does not belong to this quasar population. 
Therefore, we will not expect to find a high value of blueshift for our source. \\
Moreover several works in the literature (e.g., Shen \& Liu 2012; Coatman et al. 2016) found that for small values of FWHM and $\sigma_{line}$ (<5000 km s$^{-1}$), a low value of blueshift ($\Delta \rm v$<2000 km~s$^{-1}$) is usually observed.\\
We computed the C$\rm IV$ line blueshift of PSO~J0309+27 following the equation of Coatman et al. (2017): $\Delta$${\rm v~(km~s^{-1})}$ = $c\frac{1549.48\AA-\lambda_{half}}{1549.48\AA}$, where $c$ is the speed of light, 1549.48$\mbox{\AA}$ is the rest frame wavelength for the C$\rm IV$ and $\lambda_{half}$ is the line centroid\footnote{The line centroid is defined as the wavelength that bisect the line in two equal part. We used the definition of Dalla Bont{\`a} et al. (2020): $\lambda_{half}$ = $\frac{\int \lambda P(\lambda) d\lambda}{\int P(\lambda) d\lambda}$, where P$(\lambda)$ is the line profile.}. 
The value of the estimated $\lambda_{half}$ and $\Delta \rm v$ is reported in Table~\ref{param}.
The blueshift value is smaller than 2000 km~s$^{-1}$, as we expected, which is an indication that the outflows component is weak with respect to the emission of virialized gas. 
This allows us to infer that the virial black hole mass of PSO~J0309+27 computed in Sect.~\ref{vir} should not be strongly affected by blueshift effects.

\subsection{Object magnitude in J and K bands}
Once measured the ZP thanks to the photometric stars, we derived the $J$ and $K'$ magnitude of PSO~J0309+27 by using the IRAF aperture photometry package $qphot$ and the extinction curve of the observational site.
The aperture size was chosen to be three times as large as the seeing.
$J$ and $K'$ AB magnitude of PSO~J0309+27 are reported in Table \ref{tngtable}.
The conversion factor (in Vega system) to switch from $K'$ to $K$ magnitude has been computed by convolving $K$ and $K'$ NICS filter transmission curves with an A0 stellar template. 
We found: $K'$ = $K$ + 0.159. 
Therefore, we obtained: $K$ (AB) = 20.77 $\pm$ 0.08.
The extracted $J$ magnitude is consistent with that found for PSO~J0309+27 in the UKIRT Hemisphere Survey (UHS, Dye et al. 2018): 19.51 $\pm$ 0.38 (Vega) = 20.42 $\pm$ 0.38 (AB).  

\subsection{Black hole mass estimation}
\label{mass}
We computed the central black hole mass (M$_{\rm BH}$) of PSO~J0309+27 following two different and independent methods. 
The first method is the commonly used virial approach (the single epoch, SE, method) and the second is based on the modeling of the accretion disk emission.

\subsubsection{Single epoch mass}
\label{vir}
The SE approach is the most used and reliable method to compute black hole masses of Type I un-obscured AGNs. 
Although some works have questioned the reliability of C$\rm IV$ as a good virial mass indicator (e.g., Sulentic et al. 2007; Shen \& Liu 2012; Trakhtenbrot \& Netzer 2012) due to its observed blueward asymmetry and velocity shifts of the line profile, other authors have demonstrated that there are no large inconsistencies between the SE M$_{\rm BH}$ computed from C$\rm IV$ and Balmer lines (e.g., Vestergaard \& Peterson 2006; Greene et al. 2010; Assef et al. 2011, Dalla Bont{\`a} et al. 2020). \\
To compute the black hole mass of PSO~J0309+27 we followed the scaling relation of Vestergaard \& Peterson (2006) based on the $\sigma_{line}$\footnote{In Appendix \ref{massfwhm} we reported the M$_{\rm BH}$ derived by using the FWHM, to better facilitate the comparison with SMBH masses of other high-z AGNs in the literature.}:
\begin{equation}
\label{masseq}
M_{\rm BH} = 10^{6.73} \times (\frac{\sigma (km/s)}{10^3 km/s})^2 \times (\frac{\lambda L_{\lambda_{1350\AA}}}{10^{44} erg/s})^{0.53}
\end{equation}
We used this relation because Denney et al. (2013) and Dalla Bont{\`a} et al. (2020) find better agreement between C$\rm IV$-based and H$_{\beta}$-based mass estimates by using $\sigma_{line}$ rather than FWHM, in particular when high quality spectra are used.
Moreover we did not use black hole mass estimators that correct the effect of the C$\rm IV$ line blueshift (e.g., Coatman et al. 2017) for the following reasons: 1) the $\sigma_{line}$ parameter is relatively insensitive to the blueshift\footnote{The blueshift correction for black hole mass estimators has been calibrated only for the FWHM parameter (e.g., Coatman et al. 2017).} (e.g., Coatman et al. 2017, Dalla Bont{\`a} et al. 2020); 2) Coatman et al. (2017) suggest to use these estimators when the blueshift value is larger than 3000 km s$^{-1}$ (the values measured for PSO~J0309+27 is not so high, see Sect. \ref{shift}); 3) the application of the blueshift correction factor, calibrated on z$<$4 AGNs may be inappropriate for sources at higher redshifts (e.g., Park et al. 2017; Mej{\'\i}a-Restrepo et al. 2018; Kim et al. 2018).
The continuum luminosity at 1350$\mbox{\AA}$ ($\lambda L\lambda_{1350\AA}$) has been computed directly from the PS1 $y$ point (rest-frame wavelength = $\sim$1370$\mbox{\AA}$):  
$\lambda L_{\lambda_{1350\AA}}$ = 2.49$\pm$0.32$\times$10$^{46}$ erg s$^{-1}$.\\
Therefore, from Eq. \ref{masseq} we obtained a M$_{\rm BH}$ equal to 1.45$^{+0.25}_{-0.22} \times 10^9$ M$_{\odot}$.
The reported uncertainty, derived by propagating the errors of the C$\rm IV$ line width and of the monochromatic continuum luminosity, is purely statistical.
By taking into account the large intrinsic scatter of the C$\rm IV$ relation of the SE method ($\sim$0.36~dex, e.g., Vestergaard \& Peterson 2006; Denney et al. 2012; Jun et al. 2017), that dominates the overall M$_{\rm BH}$ error, we obtained a black hole mass of 1.45$^{+1.89}_{-0.85}\times$10$^9$M$_{\odot}$.\\
Since PSO~J0309+27 is an object observed under a small viewing angle, it is important to take into account the potential problematics related to the use of the SE method on this type of source.
First, since the Broad Line Region (BLR) may not be isotropic the resulting black hole mass could be systematically underestimated in objects observed face-on. 
Indeed, there is a large consensus in the literature (e.g., McLure \& Dunlop 2002; Decarli et al. 2008, 2011) concerning to a disk-like structure of the BLR. 
However, it is not already clear if the width of the broad emission lines (including C$\rm IV$) depends on the orientation. 
For instance, Runnoe et al. (2014, and reference therein) found this dependence in RL AGNs for H$\beta$, but not for the C$\rm IV$ line. Similarly, Fine et al. (2011) in a sample of RL AGNs did not find a correlation between the line width of the C$\rm IV$ and the AGN orientation. 
These authors concluded that the high ionization lines are produced in isotropic inner parts of the BLR.\\
A second potential issue connected with the orientation of the source is the fact that the AGNs continuum luminosity could be contaminated by the relativistic jet. 
Therefore, a continuum-luminosity based relationship may lead to a mass overestimate (e.g., Decarli et al. 2011). 
Moreover, we also have to take into account the possible anisotropy of the continuum emission produced by the accretion disk.
Therefore, the observed continuum luminosity is higher for a source viewed face-on (e.g., Calderone et al. 2013). 
This effect could lead to overestimate mass, since the SE relations are empirically calibrated on type-I AGNs randomly oriented (with an expected mean angle of $\sim$30$^{\circ}$). 
However, there are no evidence of the presence of these potential bias for oriented RL AGNs, as it has been recently demonstrated by Diana et al. (2021, MNRAS, submitted). 
In this work the authors did not find a significant difference between the ratio between the C$\rm IV$ line luminosity (which is not affected by the beaming) and the continuum luminosity at 1350$\mbox{\AA}$ (which could be affected by the beaming) of a sample of $\sim$380 blazars with that of the RQ AGNs of the sample of Shen et al. (2011), for which the beaming is not present and that are, on average, observed at different angle compared to blazars. 
We computed this luminosity ratio (R=L$_{1350\AA}$/L$_{C\rm IV}$) also for PSO~J0309+27, finding that it is at 1$\sigma$ from the mean value of the sample of Shen et al. (2011).
This allows us to infer that the peculiar orientation of PSO~J0309+27 does not affect the observed continuum emission and, hence, the derived SE black hole mass.

\subsubsection{Accretion disk model}
\label{ad}
\begin{figure}[!h]
	\centering
	\includegraphics[width=9.cm, height=7.0cm]{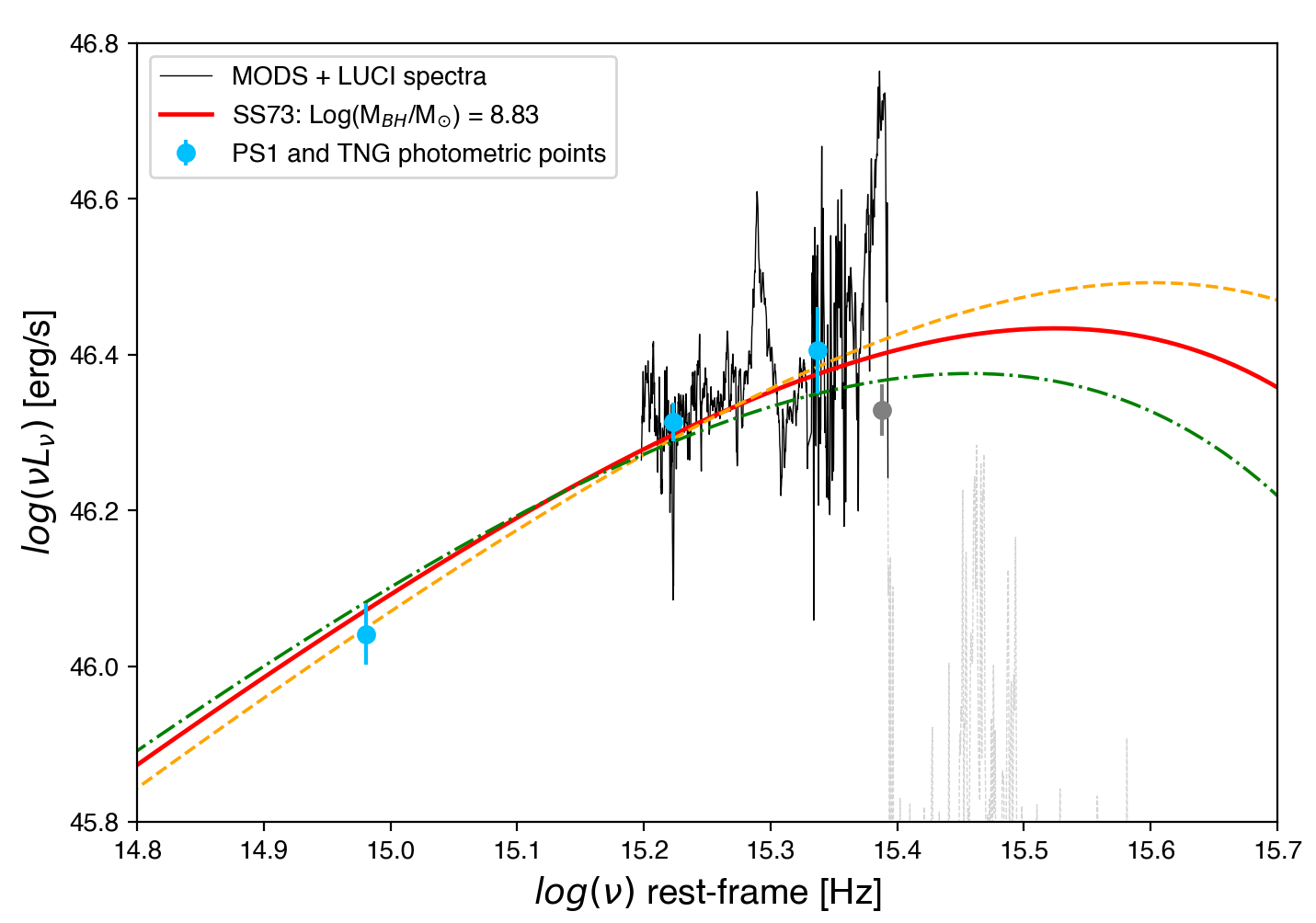}
    \caption{\small Accretion disk model of PSO~J0309+27. The SS73 model that best represents our data is shown in red. Orange dashed line and green dashed-dotted line represent the model with the minimun and maximun value of M$_{\rm BH}$ respectively (5.2$\times$10$^8$ and 8.8$\times$10$^8$ M$_{\odot}$). 
   In black the optical and NIR (smoothed) spectra are reported. 
The PS1 $y$ and the TNG $J$ and $K$ photometric data are shown in light blue. In grey we highlighted the photometric point and the part of the spectrum 
   affected by HI absorption (therefore not used for the disk modeling).}
	\label{mass_fit}	
\end{figure}

To verify the presence of any possible bias on the calculated SE masses we used an independent technique based on the accretion disk emission (e.g., Sbarrato et al. 2012; Calderone et al. 2013; Ghisellini et al. 2015; Belladitta et al. 2019; Paliya et al. 2020; Diana et al. 2021, MNRAS, submitted). 
This technique assumes that the optical/UV continuum emission of the AGN is produced by an optically thick, geometrically thin accretion disk (AD) that emits according to the Shakura \& Sunyaev (1973, SS73) model.
The SS73 assumes a non--spinning\footnote{Usually RL AGNs are associated to spinning black holes. However the assumption of a non-spinning black hole is justified also in the case of PSO~J0309+27 by the results of Campitiello et al. (2018) who found an equivalence between the accretion disk fit with a SS73 model and a KerrBB model with spin$\sim$0.8 observed face-on (as expected for blazars).} black hole (i.e., the efficiency of the accretion process, $\eta$, is $\sim$0.1), surrounded by an AD divided in rings that emit as black bodies.
The total disk luminosity is therefore a superposition of black body spectra with the form:
\begin{equation}
\label{L}
L(\nu, M_{BH}, \dot{M})d\nu = 4\pi^2 \int_{R_{in}}^{R_{out}} RB_{\nu}[T(R,M_{BH}, \dot{M})]d\nu dR
\end{equation}   
where $\dot{M}$ is the mass accretion rate, R is the distance from the central engine and B$_{\nu}[T(R,M_{BH}, \dot{M})]d\nu$ is the Planck's spectrum.
The temperature of the single black body emission depends on the distance from the central black hole, with the highest temperature emitted at the disk inner radius, which corresponds to the innermost stable circular orbit: $R_{\rm in}=3R_{\rm Schw}=6GM_{\rm BH}/c^2$ (e.g., Misner et al. 2017 and reference therein).
A large black hole mass implies a larger disk inner radius and hence a lower emitting temperature (as T $\propto$ M$_{\rm BH}^{-1/2}$). 
Therefore the accretion disk radiation peaks at lower frequency compared to smaller masses.\\
With these assumptions it is possible to derive the values of M$_{\rm_{BH}}$ and of the accretion rate ($\dot{M}$), that are free parameters of the SS73 model, simply by fitting the optical/UV data points that do not suffer from HI absorption, i.e. those at frequency lower than the Ly-$\alpha$ line.
With the main goal of testing the SE masses, we decided to apply the AD method to PSO~J0309+27 even if the optical/UV spectral range is not well sampled (see Fig.~\ref{mass_fit}), since we have only three photometric points not affected by HI absorption: the $y$ data from PS1 and the data at $J$ and $K$ bands from the TNG follow-up.
In addition, the continuum of the LBT spectrum has been used as a guide line for the modeling. 
By using this method we found that PSO~J0309+27 can be described by a black hole of M$_{\rm BH}$ = 6.9$_{-3.7}^{+1.9}$$\times$10$^8$ M$_{\odot}$. 
To reduce the uncertainty interval on the estimated mass we have used the expected value of the peak luminosity (L$_{peak}$) of the accretion disk as a further constraint. L$_{peak}$ can be inferred from L$_{C\rm IV}$ and/or from $\lambda L_{\lambda_{1350\AA}}$ by using the relations found in Calderone et al. (2013, equations (2) and (5) of this paper) and by taking into account the inclination of the source (i.e, the fact that PSO J0309+27 is a blazar, $\theta \sim$ 0$^{\circ}$). 
This constraint allowed us to infer that the value of the black hole mass of the source cannot be lower than 5.2$\times$10$^8$ M$_{\odot}$. 
Therefore the final black hole mass of PSO J0309+27 computed with the accretion disk method is: M$_{\rm BH}$ = 6.9$_{-1.7}^{+1.9}$$\times$10$^8$ M$_{\odot}$ (see Fig.~\ref{mass_fit}).
This value is consistent with that obtained from the virial method, considering the total uncertainties.

\subsection{Bolometric Luminosity and Eddington ratio}
\label{edd}
Using the value of the SE black hole mass, we derived the Eddington luminosity (L$_{Edd}$) and the Eddington ratio ($\lambda_{Edd}$). The latter quantifies how fast the accretion rate is with respect to the Eddington limit.
To compute $\lambda_{Edd}$ we first estimated the bolometric luminosity (L$_{bol}$) of PSO~J0309+27 using a bolometric correction (e.g. Richards et al. 2006): L$_{bol}$ = L$_{1350\AA}$ $\times$ K$_{bol}$.
In this case we used the bolometric correction factor of Shen et al. (2008): K$_{bol}$ = 3.81$\pm$1.26. 
However we have to remind that this K$_{bol}$ is calibrated empirically over RL and RQ Type I AGNs, with a mean expected angle of 30$^{\circ}$. 
Since PSO~J0309+27 is a source seen under a small viewing angle ($\theta \sim$ 0$^{\circ}$) and that the continuum emission from the disk is not isotropic, as already mentioned in Sect. \ref{vir}, we have to take into account the expected inclination factor ($i$=$\frac{\cos 0^{\circ}}{\cos 30^{\circ}}$ = 1.15) to compute the intrinsic bolometric luminosity. 
This led to a final estimate of: L$_{bol}$ = 8.22$\pm$3.70$\times$10$^{46}$ erg~s$^{-1}$.
We obtained a similar value for L$_{bol}$ ($\sim$8$\times$10$^{46}$ erg~s$^{-1}$) by using the non linear relation between L$_{bol}$ and L$_{1350\AA}$ of Runnoe et al. (2012).
Then we computed $\lambda_{Edd}$ as as the ratio between the bolometric luminosity (i.e. that includes the optical/UV radiation of the accretion disk, the emission reprocessed by the molecular torus and the X---ray corona radiation) and the Eddington luminosity derived from both the $\sigma_{line}$ and the FWHM\footnote{See Appendix~\ref{fwhm} for details.}. 
In the following sections we always use the $\lambda_{Edd}$ estimated from the $\sigma_{line}$, which we considered the best BH mass estimator. 
The obtained value is: $\lambda_{Edd}$ = 0.44$^{+0.78}_{-0.35}$. 
The uncertainty already takes into consideration both the statistical error on the virial mass and the intrinsic scatter of the SE relation ($\sim$0.36 dex).\\
The values of the SE M$_{\rm BH}$ and of $\lambda_{Edd}$ of PSO~J0309+27 are in line with those derived for RL and RQ AGNs discovered at similar redshift (z=5.5-6.5)\footnote{SE black hole masses of 5.5$\leq$z$\leq$6.5 RQ and RL AGNs reported in the literature varies from 8$\times$10$^{7}$ M$_{\odot}$ to 1$\times$10$^{10}$M$_{\odot}$; the value of $\lambda_{Edd}$ varies from 0.03 to 1.3; Jiang et al. 2007; Willott et al. 2010b; De Rosa et al. 2011; Mazzucchelli et al. 2017; Eilers et al. 2018; Kim et al. 2018; Shen et al. 2019; Onoue et al. 2019; Andika et al. 2020.}.
Broadly speaking, PSO~J0309+27 is fully consistent with a typical z$\sim$6 AGN and shows no evidence of peculiarities associated with its relativistic beamed jet.
However the similarity of masses and Eddington ratio could be likely a consequence of a selection bias, as all these high-z sources have been selected from similar optical/IR surveys.\\
If we consider, for the $\lambda_{Edd}$ computation, only the luminosity of the accretion disk, which is L$_{disk}$ $\sim L_{bol}/2$ (e.g., Calderone et al. 2013), we obtain an Eddington ratio equal to 
0.22$^{+0.40}_{-0.19}$.

\section{Implication on the early SMBH growth}
\label{psogrowth}
\begin{figure}[!h]
	\centering
	\includegraphics[width=9cm]{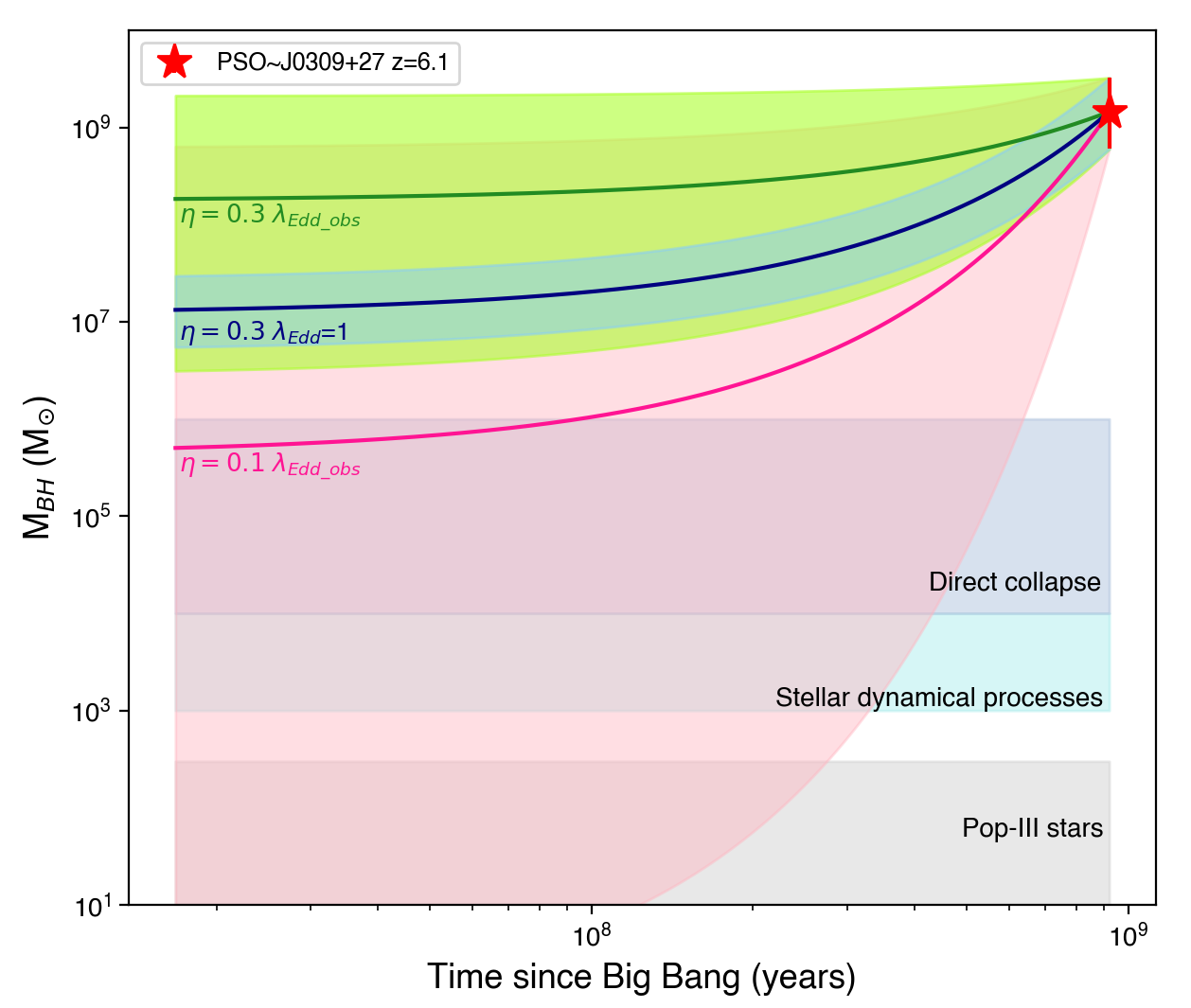}
	\caption{\small Estimated growth history of PSO~J0309+27 (red star). Solid lines represent the best fit cases, under which the corresponding $\lambda_{\rm Edd}$ and $\eta$ are reported. 
    The shaded horizontal regions correspond to the expected mass ranges of Pop III remnants BHs (M$_{seed}$$\leqslant$10$^2$M$_{\odot}$, grey), stellar dynamical processes (M$_{seed}$$\geqslant$10$^3$-10$^{4}$M$_{\odot}$, cyan) and direct collapse BHs (M$_{seed}$$\sim$10$^4$-10$^6$M$_{\odot}$, dark cyan).
	The values of the different seed black holes are taken from	Valiante et al. (2016).}
	\label{growth}	
\end{figure}

Accurate measurements of black hole masses and Eddington ratios of high-z AGNs help us constraining formation scenarios of the first seed black holes. 
Moreover, high-z RL AGNs provide a unique opportunity to study the role of jets in the accretion of early SMBHs (e.g., Volonteri et al. 2015).
In particular, as already mentioned, the discovery of jetted AGNs in the early Universe ($z>5$) represent a serious challenge to our understanding of black hole growth, especially if the presence of the jet is associated with a rapidly spinning black hole, which is expected to have a large radiation efficiency and, therefore, a longer growth time with respect to black holes hosted by RQ AGNs.\\ 
Assuming that the black hole seed grows at a constant Eddington ratio during the entire accretion process (e.g., Shapiro 2005; Volonteri \& Rees 2005), the evolution of the M$_{\rm BH}$ with time is directly proportional to the mass itself, resulting in an exponential growth from the initial mass (M$_{BH, seed}$):
\begin{equation}
\label{tgrow}
	M_{BH, seed} = M_{BH} \times exp(-\frac{t_{growth}}{\tau})
\end{equation}
where t$_{growth}$ is the time during which the black hole accretes and $\tau$ is the e-folding timescale:
\begin{equation}
	\tau = 0.45 (\frac{\eta}{1-\eta})(\frac{1}{\lambda_{\rm Edd}})(\frac{1}{f_{act}(M, t)}) \hspace{0.1cm} Gyr
\end{equation}
where $f_{act}$ is the duty cycle of the black hole, i.e. the mass and redshift dependent fraction of time when the black hole is active ($f_{act}(M,t)$=[0,1]).
Therefore, from Eq. \ref{tgrow} we can derive the initial mass of the black hole seed required to observe the mass of PSO~J0309+27 at z$\sim$6 (0.922~Gyr after the Big Bang).
Figure \ref{growth} shows the estimated growth history of PSO~J0309+27 according to this model.
These results depend on the assumptions made, i.e. the redshift of the seed formation, the accretion rate, the radiative efficiency, the value of $f_{act}$.
We traced the mass back to z=30, when the first stars and galaxies are thought
to have formed (e.g., Bromm \& Larson 2004; Bromm \& Yoshida 2011).
The value of the duty cycle ($f_{act}$) is assumed to be equal to 1, meaning that the AGN has been active for all the time.
Then we considered different values for $\eta$ and $\lambda_{\rm Edd}$. 
We assumed that the seed black hole accretes constantly with the observed Eddington ratio ($\lambda_{\rm Edd}$ = 0.44) or with the value of 1, the maximum value allowed in an Eddington limited accretion scenario.
The efficiency parameter, instead, is believed to depend on black hole spin (e.g., King \& Pringle 2006) and can be as high as $\sim$30-40\% in case of spinning black hole (e.g., Thorne 1974; Reynolds 2014).  
Current semi-analytical models place only weak constraints on the spin values for AGNs at z$>$5, which depend on the gas accretion mode, the morphology of the host galaxy and black hole mass (e.g., Sesana et al. 2014). 
Therefore, as no stringent constraints on black hole spin are reported to date for high-z AGNs, we assumed both an efficiency of 0.1 (typical of not rapidly spinning black holes) and 0.3 (typical of Kerr black holes).
Figure \ref{growth} shows that only a scenario of $\eta$=0.1 can reproduce a theoretically accepted seed mass.
Scenarios of higher efficiency ($\eta$=0.3), instead, would require more massive seeds (M$_{seed}$$\geq$10$^6$M$_{\odot}$) as progenitors of PSO~J0309+27. 
These expected seeds are even more massive than what direct collapse models predict (M$_{seed}$$\sim$10$^4$-10$^6$M$_{\odot}$, Latif \& Ferrara 2016).\\
This result suggests that such high values of efficiency are probably not realistic, not even for RL AGNs.
Alternatively, super-Eddington accretion episodes must occur for a significant fraction of the growth time. 
To date, there have been no clear examples of such super-Eddington SMBHs at z$>$6, although this scenario has been suggested for J1205$-$0000, a mildly obscured AGN at z=6.699 (Onoue et al. 2019), for PSO~J006+39 at z=6.621 (Tang et al. 2019), two RQ AGNs, and for PSO~J172+18, the most distant RL AGN ever discovered (z=6.8, Ba{\~n}ados et al. 2021).
Super-Eddington accretion episodes are often taken into consideration also for the growth of black holes hosted in the z$>$7 RQ AGNs discovered so far (e.g., Bañados et al. 2018a; Wang et al. 2021).  
It has been suggested that maintaining super-Eddington accretion might be possible in specific environments (e.g., dust-obscured AGNs with strong winds or gas rich AGNs; Kim et al. 2015; Kubota \& Done 2019; Moffat 2020), but whether or not this type of accretion is sustainable remains an important open question for the growth of both RL and RQ AGNs.\\
Another possible solution taken into consideration for the growth of black holes in RL AGNs has been proposed by Jolley \& Kunzic (2008), Jolley et al. (2009) and Ghisellini et al. (2010a). 
These authors proposed that when a jet is present, not all the gravitational energy of the infalling matter is transformed into heat and then radiation, but can be transformed in other forms, such as amplifying the magnetic field energy of the inner disk, a necessary ingredient for launching the jet (Blandford \& Znajek 1977).
In this case the total efficiency of the accretion process can be equal to 0.3, but only a fraction of it ($\eta_d$, i.e. the radiation efficiency) goes to heat the disk, while the rest (1$-\eta_d$) amplifies the magnetic field necessary to launch the jet.
Therefore, disk luminosity becomes Eddington limited for a larger accretion rate, making the black hole growing faster. \\
Cosmological simulations of seed black holes growth (e.g., Di Matteo et al. 2008; Alexander \& Hickox 2012; Feng et al. 2014a) are fundamental to understand what are the main ingredients of the black hole seed evolution besides the accretion process. 
In particular, mergers and AGN feedback could be taken into account for understanding the evolution of SMBHs hosted in RL AGNs. 
Indeed, RL AGNs are commonly found in rich environments at different cosmic epochs (from redshift 0.5 to z=5.8; e.g., Pentericci et al. 2000; Venemans et al. 2002, 2004; Zheng et al. 2006; Hatch et al. 2014). 
Theoretical models strongly support a preferential over-dense environment around RL AGNs (e.g., Orsi et al. 2016; Izquierdo-Villalba et al. 2018) and similar conclusions have been also found by studying the RL AGNs level of clustering with cosmic times (e.g., Magliocchetti et al. 2004; Retana-Montenegro \& R{\"o}ttgering 2017). 
All these results suggest that the presence of a relativistic jet may indeed be preferentially triggered in dense environments (i.e., in \emph{protoclusters}), where
frequent mergers between star-forming galaxies help to increase the mass and spin of the SMBHs (e.g., Hatch et al. 2014). 
In this context a study of the environment of PSO~J0309+27 could be crucial to better understand its black hole growth.

\section{Summary and conclusion}
\label{conc}
In this paper we have reported new photometric and spectroscopic observations in the NIR band of of PSO~J0309+27, the most distant blazar discovered thus far. 
From a LUCI/LBT spectroscopic observation we have detected the C$\rm IV \lambda1549$ broad emission line, which allowed us to compute the mass of the SMBH hosted by the source.
By parameterizing the C$\rm IV$ line width with the $\sigma_{line}$ and by using the SE method we estimated a mass for the central SMBH of 1.45$^{+1.89}_{-0.85}$$\times$10$^9$ M$_{\odot}$. 
Moreover, thanks to a dedicated follow-up with the TNG in $J$ and $K'$ bands we also better constrain the NIR SED of the source, allowing us to derive and independent estimate of the SMBH mass using a method based on the accretion disk emission. 
The agreement of the two results supports the reliability of our estimate. \\
The value of the black hole mass, the Eddington ratio and the bolometric luminosity of PSO~J0309+27 are in line with those of other RQ and RL AGNs at similar redshift.
However, to fully understand if the high-z blazar population is different from the RQ and/or RL ones in term of black hole mass and $\lambda_{Edd}$, a larger and statistically complete sample of blazars at the highest redshift is required. \\ 
Finally, we have computed the mass of the seed black hole required to reproduce the mass of the SMBH hosted by PSO~J0309+27, using a simple model for the SMBH growth. 
We found that to obtain a reasonable (i.e., predicted by the models) seed black hole the efficiency of the accretion process can not be as high as 0.3, as expected for a SMBH hosted by a RL AGN. 
A high efficiency of 0.3 could be possible if super-Eddington accretion episodes are taken into account during the black hole growth or only a part of the released gravitational energy of the infalling matter is used to heat the accretion disk.\\ 
Future studies on the environment of PSO~J0309+27 will be useful to better understand its growth and evolution.

\begin{acknowledgements}
We thank the referee for his/her useful comments that improved the quality of the manuscript.
This work is based on observations made with the Large Binocular Telescope (LBT, program DDT\_2019B\_3). We are grateful to the LBT staff for providing the observations for this object.
LBT is an international collaboration among institutions in the United States of America, Italy, and Germany. The LBT Corporation partners are the University of Arizona on behalf of the Arizona university system and the Istituto Nazionale di Astrofisica.
This work is based on observations made with the Italian Telescopio Nazionale Galileo (TNG, program A42DDT4) operated on the island of La Palma by the Fundación Galileo Galilei of the INAF (Istituto Nazionale di Astrofisica) at the Spanish Observatorio del Roque de los Muchachos of the Instituto de Astrofisica de Canarias.
We acknowledge financial contribution from the agreement ASI-INAF n. I/037/12/0 and n.2017-14-H.0 and from INAF under PRIN SKA/CTA FORECaST.
CS acknowledges financial support from the Italian Ministry of University and Research - Project Proposal CIR01\_00010.
AR acknowledges support from the INAF project Premiale Supporto Arizona \& Italia.
This research made use of Astropy (http://www.astropy.org) a community developed core Python package for Astronomy (Astropy Collaboration et al. 2018).
\end{acknowledgements}

\appendix
\section{PSO~J0309+27 black hole mass and parameters computed from FWHM}
\label{fwhm}
In this work we have used the SMBH mass based on the $\sigma_{line}$ parameter as the best M$_{\rm BH}$ estimator. 
Here we report the black hole mass, and the parameter related to it, computed from the FWHM, for a direct comparison with the literature. \\
The single epoch scaling relation of Vestergaard \& Peterson (2006) based on the FWHM is:
\begin{equation}
\label{massfwhm}
M_{\rm BH} = 10^{6.66} \times (\frac{FWHM (km/s)}{10^3 km/s})^2 \times (\frac{\lambda L\lambda_{1350\AA}}{10^{44} erg/s})^{0.53}
\end{equation}
The value of the FWHM of the C$\rm IV$ line has been computed in Sect. \ref{linewidth} and it is reported in Tab. \ref{param}. 
The luminosity of the AGN continuum is reported in Sect. \ref{mass}.
From \ref{massfwhm} we derived a black hole mass of 7.47$^{+6.93}_{-3.63}$$\times$10$^{8}$M$_{\odot}$, that is consistent with that obtained by $\sigma_{line}$ and with the AD method. 
Also on this estimate we have to take into account an intrinsic scatter of $\sim$0.4~dex. 
Since the value of the blueshift of PSO~J0309+27 is smaller than 3000~km~s$^{-1}$ (see Sect.~\ref{linewidth}), by using the equation of Coatman et al. (2017) that correct the virial mass for the blueshift effect, we did not expect to find a significantly different value. Indeed we obtained: M$_{\rm BH\_corr}$ = 5.80$^{+5.48}_{-3.53}\times10^8$M$_{\odot}$, consistent with the previous one.
From the virial black hole mass and by using the same value of bolometric luminosity reported in Sect.~\ref{edd} we computed an Eddington ratio of  0.86$^{+1.20}_{-0.80}$, which is consistent with that computed from the $\sigma_{line}$.

\end{document}